\documentclass[structabstract]{aa}  
\usepackage{graphicx}
\usepackage{txfonts}

\begin{document}
   \title{An astroclimatological study of candidate sites to host an imaging atmospheric Cherenkov telescope in Romania}

   \titlerunning{An astroclimatological study for a Cherenkov telescope in Romania}

   \author{A. A. Radu
          \inst{1}
          \and
		   T. Angelescu
		   \inst{2}
		   \and
		   V. Curtef
		   \inst{3}
		   \and
		   D. Felea
		   \inst{1}
		   \and
		   D. Hasegan
		   \inst{1}
		   \and
		   B. Lucaschi
		   \inst{4}
		   \and
		   A. Manea
		   \inst{4}
		   \and
		   V. Popa
		   \inst{1}
		   \and
		   I. Ralita
		   \inst{4}
		   }

   \institute{Institute for Space Sciences (ISS), 409 Atomistilor St., Bucharest-Magurele, Romania\\
              \email{aurelian.radu@spacescience.ro}
              \and
              Bucharest University, Faculty of Physics, P.O. BOX MG-012, Bucharest-Magurele, Romania
			  \and
			  Universit$\mathrm{\ddot{a}}$t W$\mathrm{\ddot{u}}$rzburg, D-97074 W$\mathrm{\ddot{u}}$rzburg, Germany
			  \and
			  National Meteorological Administration, 97 Soseaua Bucuresti-Ploiesti, Bucharest, Romania
             }


  \abstract
   {We come out in this paper with an astroclimatological study of meteorological data on relative humidity, dew point temperature, air temperature, wind speed, barometric air pressure, and sky cloudiness recorded at four Romanian locations (Baisoara, Rosia Montana, Semenic, Ceahlau) and Nordic Optical Telescope (NOT) located at Observatorio del Roque de Los Muchachos, in the Canary Islands.}
   {Long term trends of microclimates have been compared in order to identify the site-to-site variations. We have performed this analysis as part of a site testing campaign aimed at finding the best location for the establishment of a small Cherenkov telescope in Romania. The conditions at the Romanian sites have been compared to those of the Canary Islands considered as a reference.}
   {A statistical approach has been used for data analysis. Monthly and annual samples have been extracted from series of raw data for nighttime, daytime and entire day intervals. Percentage distributions of meteorological parameters, whose values exceed certain limits, have been derived.}
   {Significant differences have been found between the Romanian sites and the NOT site. The comparison of the Romanian locations indicates Baisoara to be the best site for the establishment of the telescope, closely followed by Rosia Montana. As these two sites are both located in the Apuseni Mountains, we consider this area as the optimal place for performing astronomical observations in Romania.}
   {}


   \maketitle
%

\section{Introduction}\label{1}

The very high-energy (VHE) $\gamma$-rays (100 GeV $<$ E $<$ 100 TeV), of cosmic origin, are best detected by the ground based Imaging Atmospheric Cherenkov Telescopes (IACTs) (Buckley et al. \cite{buckley}). MAGIC (Baixeras et al.  \cite{baixeras}), VERITAS (Weekes et al. \cite{weekes}), H.E.S.S. (Hinton et al. \cite{hinton}), and CANGAROO-III (Kubo et al. \cite{kubo}), the most important IACT facilities currently in operation have reported scientific results with a huge impact on astrophysics, cosmology and particle physics. The future Cherenkov Telescope Array (CTA) (Hofmann $\&$ Martinez \cite{hofmann}) with its next generation, highly automated IACT telescopes will open a new era of outstanding precision for gamma-ray astrophysics.

All major IACT experiments perform observations at their sensitivity limit or in multi-wavelength campaigns for most of their operation time. If monitoring observations take place, the amount of time assigned is, by far, not sufficient. Under these circumstances and considering the importance of the observational data, a global network of several small Cherenkov telescopes was proposed to be operated in a coordinated way for long-term monitoring of the brightest blazars - the Dedicated Worldwide AGN Research Facility (DWARF) (Bretz et al. \cite{bretz}). FACT (First G-APD Cherenkov Telescope) (Backes et al. \cite{backes}), the prototype telescope of this network has been prepared for operations at the Roque de los Muchachos on the Canary Island of La Palma.

To join the international efforts on understanding the physics of VHE $\gamma$-rays, a small, ground based, imaging atmospheric Cherenkov telescope of last generation is intended to be installed and operated in Romania as a component of the DWARF network.

A site assessment campaign has been initiated for the identification of the best location where to establish the telescope. Technical, economical and social selection criteria were applied on a list of possible sites and four candidate locations have been selected (Baisoara, Rosia Montana, Semenic, Ceahlau) (Radu et al. \cite{radu}). An astroclimatological study based on a statistical data analysis of local meteorological parameters has been performed for these locations and the results are presented in this paper. The local microclimates at the Romanian candidate sites have been compared to that of Nordic Optical Telescope (NOT) operated on the island of La Palma (Canary Islands, Spain) (Djupvik $\&$ Andersen \cite{djupvik}).

The international scientific community has sustained over the years large efforts to find the best sites for the operation of astronomical telescopes, but now it is for the first time that an analysis with impact on the operation of a ground based Cherenkov telescope has been performed in Romania. 

The layout of the paper is as follows: in $\S$\ref{2} we discuss the influence of the atmospheric conditions on the operation of Cherenkov telescopes; in $\S$\ref{3} the statistical characteristics of the data used for analysis are presented; in $\S$\ref{4} we include the astroclimatological results of this study.


\section{Influence of the atmospheric conditions on the operation of Cherenkov telescopes}\label{2}

As for most other types of astronomy, good atmospheric conditions are extremely important for the astronomical studies performed on very high energy gamma-rays with Cherenkov telescopes. The atmospheric conditions generally affect the quality of astronomical observations as well as the safety of the instruments and they are quantitatively evaluated by meteorological parameters. The observers monitor these parameters on a continual basis in order to ensure a correct treatment of the astronomical data and the safe operation of telescopes (Cogan \cite{cogan}).
 
In the specific case of Cherenkov telescopes, it was pointed out that atmospheric changes on relative humidity, air temperature, sky cloudiness and the influence of haze and calima\footnote{fine dust or sand carried by winds from the Sahara Desert}, usually not taken into account in simulations, may lead to a systematic underestimation of the shower energy by $\sim$ 15$\%$ (Mazin \cite{mazin}).

As a function of the type of observations performed and the specific design of each telescope, the staff enforces safety limit values for the relevant meteorological parameters beyond which the telescope cannot be operated anymore. Poor weather is generally considered as the major cause of the lost observing time. In comparison, the telescope failures and the engineering account for a much smaller fraction of the lost time. This fraction was reported to be only $\sim$ 3$\%$ in the case of VERITAS (Grube \cite{grube}). 

In the following subsections we review the influence of the most significant meteorological parameters on the operation of Cherenkov telescopes.

\subsection{Relative humidity and dew point temperature}\label{2.1}

The relative humidity and the dew point temperature set the occurrence of moist and water condensation on the coldest parts of astronomical instrumentation. If dew point temperature equals air temperature and if air cools, condensation appears. However, observing is dangerous for instrumentation even though condensation has not actually shown up, but there is a high risk that this phenomenon will occur. This happens if the difference between the air temperature and the dew point temperature is smaller than a variable upper limit ($1\ ^{\circ}$C or $5\ ^{\circ}$C, Lombardi et al. \cite{lombardi}).

As a result of condensation the light sensors and the electronic equipment become wet and can be damaged. In case the condensation occurs on the optical surfaces which include the upper surface of the telescope reflector and the interior surfaces of the light concentrators, the image of the observed astronomical object degrades.   
In what concerns the relative humidity, a value of 90$\%$ was used by the NOT collaboration, by Lombardi et al. (\cite{lombardi2}), and by Murdin (\cite{murdin}) as a threshold beyond which the observations should be stopped and the telescopes turned off. Mazin (\cite{mazin}) and Jabiri et al. (\cite{jabiri}) also considered that a relative humidity level higher than 90$\%$ makes observing unfeasible. 

High levels of relative humidity also result in an increase of air conductivity. This is important for the light detectors located in the camera of a Cherenkov telescope. Especially the photomultiplier tubes (PMTs) and their voltage supplies subjected to very high voltage during normal operation can experience electrical discharges which generate irreversible damages to the circuits (Cogan \cite{cogan}; \c{C}elik \cite{celik}; Theiling \cite{theiling}).

\subsection{Wind speed}\label{2.2}

Astronomical observations are not undertaken in windy weather due to dangers associated with the wind load on the optical reflector of the telescope, on its support structure, and on the drive subsystem (Cogan \cite{cogan}). In order to ensure that none of the subsystems of the telescope are in danger, the telescope should be turned off and brought to the parking position if the wind speed exceeds an upper safety limit. This limit is telescope dependent. A survival wind speed limit is also defined for the parking position. 

The telescopes of VERITAS have positioners\footnote{devices responsible for telescope pointing} designed to operate safely at wind speeds up to 20 mph (8.9 m $s^{-1}$/32.2 km $h^{-1}$) and to survive in the parking position to wind speeds of 100 mph (44.7 m $s^{-1}$/161.0 km $h^{-1}$) (Cogan \cite{cogan}; \c{C}elik \cite{celik}). For MAGIC, it is considered that the subsystems of the telescopes are not in danger as long as the wind speed is lower than 40 km $h^{-1}$ (11.1 m $s^{-1}$) (Mazin \cite{mazin}).

Various other safety wind speed limits are discussed in the literature. The telescopes in operation at Observatorio del Roque de Los Muchachos (La Palma, Canary Islands), are closed if the wind speed exceeds 15 m $s^{-1}$, whereas those at the site of Paranal Observatory (Atacama Desert, Chile) are closed for wind speeds in excess of 18 m $s^{-1}$ (Lombardi et al. \cite{lombardi}). In a meteorological study performed for the Oukaimeden site in Morocco, 15 m $s^{-1}$ was also indicated as a typical maximum safe operating value for the wind speed (Murdin \cite{murdin}; Jabiri et al. \cite{jabiri}). 

If high speed winds carry large amounts of dust the negative impact on the various hardware subsystems of a telescope is even worse. Besides that, dust is also a serious cause of decreasing transparency. The quantity of dust in the air depends on the altitude of the observing site, its proximity to a dust source, and the prevailing wind direction. From this point of view the Romanian sites located in areas covered with vegetation and away from dust sources are privileged.

\subsection{Air temperature}\label{2.3}

The air temperature fluctuations impact on the instrumentation of Cherenkov telescopes and on the reconstruction of shower images. 

The PMT noise increases with temperature due increased thermionic emission whereas the PMT gain reduces with temperature. Cogan (\cite{cogan}) reported a gain decrease of ($\sim$ 1$\%~$/$\ ^{\circ}$C for these sensors. The temperature inside the camera of the telescope can increase as a result of the exterior temperature or because of the heat dissipated by the electronic components located inside it (ex. preamplifiers). This is especially dangerous while performing maintenance operations during daytime in warm seasons. As a safety precaution, exhaust fans and temperature sensors are usually installed inside the telescope camera (\c{C}elik \cite{celik}; Theiling \cite{theiling}).

On the other hand, temperatures below $0\ ^{\circ}$C generate freezing. In order to avoid dew and snow effects, the mirror tiles of the large area reflector of MAGIC telescopes contain heating elements (White \cite{white}).

For Cherenkov telescopes, the main contribution to atmospheric attenuation comes from Rayleigh scattering, which depends on the chemical composition and the density of air molecules. The density of air molecules depends in its turn, on air temperature and barometric air pressure (Fruck \cite{fruck}). As atmospheric attenuation is included in the shower reconstruction algorithm, air temperature and barometric air pressure fluctuations have an impact on the flux and the energy estimates of primary gamma-rays. 

\subsection{Barometric air pressure}\label{2.4}

Generally, the air pressure analysis is performed by the astronomic community in order to see if a certain site is dominated by high pressure which would imply prevailing stable good weather (Lombardi et al. \cite{lombardi2}). 

As one moves from the sea level to higher altitudes, the air pressure gets lesser and lesser. The theoretically expected barometric air pressure ($\mathrm{P_{theo}}$) at a given altitude can be calculated according to the US Standard Atmosphere model (\cite{usstatm}) with the following formula:
\begin{equation}\label{eq1}
P=P_{b}\left[\frac{T_{b}}{T_{b}+L_{b} \cdot \left(h - h_{b}\right)}\right]^{\frac{g_{0} \cdot M_{0}}{R^{*} \cdot L_{b}}},
\end{equation}
where $\mathrm{g_{0}}$ = 9,80665 m $\mathrm{s^{-2}}$ represents the sea-level value of the acceleration of gravity, $\mathrm{M_{0}}$ = 0.0289644 kg $\mathrm{mol^{-1}}$ is the molar mass of Earth's air, and $\mathrm{R^{*}}$ = 8.31447 J $\mathrm{K^{-1}}$ $\mathrm{mol^{-1}}$ is the Universal gas constant. For altitudes lower than 11 km (troposphere), subscript b value is zero and it refers to sea-level. Then the geopotential height $\mathrm{h_{b} = 0}$, and the standard temperature lapse rate $\mathrm{L_{b} = -0.0065}$ K $\mathrm{m^{-1}}$. If we replace all these constants in Equation~(\ref{eq1}), we obtain:
\begin{equation}\label{eq2}
P=P_{0}\left(1 - 0.0065\frac{h}{T_{0}}\right)^{5.256},
\end{equation}
where $\mathrm{P_{0}}$ = 1013.25 hPa is the sea level standard atmospheric pressure, and $\mathrm{T_{0}}$ = 288.15 K ($\mathrm{15\ ^{\circ}}$C) is the sea level standard temperature. For each investigated location, we have used Equation~(\ref{eq2}) to compute $\mathrm{P_{theo}}$ and the results are included in Table~\ref{tabel1} together with the geographical positions and the altitudes. 
   \begin{table*}
   \caption{The geographical positions, the altitudes and the theoretically expected barometric air pressure ($\mathrm{P_{theo}}$) for the analyzed sites.}
   \label{tabel1}
   \centering
   \begin{tabular}{c c c c c}
   \hline\hline
   Site	& Latitude	& Longitude	& Altitude [m]	& $\mathrm{P_{theo}}$ [hPa] \\
   \hline
   Rosia Montana	& 46$^{o}$ 19$'$ 03$"$ N	& 23$^{o}$ 08$'$ 21$"$ E	& 1198	& 877.5 \\
   Baisoara	& 46$^{o}$ 32$'$ 08$"$ N	& 23$^{o}$ 18$'$ 37$"$ E	& 1357	& 860.6 \\
   Semenic	& 45$^{o}$ 10$'$ 53$"$ N	& 22$^{o}$ 03$'$ 21$"$ E	& 1432	& 852.7 \\
   Ceahlau	& 46$^{o}$ 58$'$ 39$"$ N	& 25$^{o}$ 57$'$ 00$"$ E	& 1897	& 805.1 \\
   NOT	& 28$^{o}$ 45$'$ 26$"$ N	& 17$^{o}$ 53$'$ 06$"$ W	& 2382 &	758.0 \\
   \hline
   \end{tabular}
   \end{table*}
The mean pressure value for a certain location, determined from statistical analysis, is compared to the theoretically expected air pressure ($\mathrm{P_{theo}}$) for that altitude. The frequency with which it happens for the mean pressure to be higher than $\mathrm{P_{theo}}$ is an indirect indication of high atmospheric stability (Jabiri et al. \cite{jabiri}). On the other hand, variations of barometric air pressure in a short time scale (few hours) can induce weather instabilities with a serious impact on the operation of a telescope (Lombardi et al. \cite{lombardi}). 
As the barometric air pressure may anticipate the temperature behavior by 2-3 hours (Lombardi et al. \cite{lombardi2}), the information carried by the data can be used to better implement the thermalization of the telescope and of the instruments. 
In what concerns the Cherenkov telescopes, the barometric air pressure fluctuations impact on the atmospheric attenuation and implicitly on the shower reconstruction as discussed at the end of Subsection~$\S$\ref{2.3}. 

\subsection{Sky cloudiness}\label{2.5}

The clouds positioned directly overhead or not, can modify the response of a Cherenkov telescope. They can block the light coming from one direction and reflect the light coming from another. This happens for the light generated within the extensive air showers, and also for the light originated from other sources (e.g. undesired city lights). This way either some useful light is lost or the sky may look brighter than it should be (Valc\'{a}rcel \cite{valcarcel}).

When clouds move into the field of view of a telescope they absorb Cherenkov radiation from extensive air showers and drops in the trigger rates result, rendering any background analysis or absolute flux calculations very difficult (Cogan \cite{cogan}). A trigger rate lower than expected should always be accompanied by a lower number of recognized stars provided by the starguider, in order to be sure that a cloudy atmosphere is the cause (Mazin \cite{mazin}). In case only the trigger rate decreases, but the number of the recognized stars does not, this rather points to hardware problems of the telescope. 

Over the years, the standard procedure to evaluate the sky cloudiness has implied human observers going out, next to the telescope, and judging by themselves the condition of the sky. In the case of Whipple and VERITAS experiments, the observers gave a letter grade rating for the sky conditions at the moment of observations (Valc\'{a}rcel \cite{valcarcel}; Smith \cite{smith}). Even though, the procedure can be regarded as arbitrary because clouds are hard to see at night and it is considered inconsistent and subjective due to its human grading, it may still be regarded as a useful first criterion for judging the validity of data (Smith \cite{smith}).

An improvement of this situation was realized when a far-infrared (FIR) pyrometer connected to a telescope started to be used. The device was employed to detect the high thin wispy clouds that were difficult to see at night, by continuously measuring the temperature of the night sky (\c{C}elik \cite{celik}). This type of cloud could have on data, subtle effects which were poorly understood, and could have detrimental results on the analysis of source flux. 

Other tools for monitoring sky conditions included optical cameras (Kenny \cite{kenny}) and weather satellite maps (Valc\'{a}rcel \cite{valcarcel}).

\section{Statistical characteristics of the data}\label{3}

For the present study we have sampled data from the archive of the National Meteorological Administration (NMA) of Romania. The meteorological data were collected in the period January 1, 2000 to December 31, 2009 by the NMA weather stations operated in the locations of Baisoara, Rosia Montana, Semenic and Ceahlau. The geographical positions and the altitudes of these locations are presented in Table~\ref{tabel1} and more details can be found in (Radu et al. \cite{radu}). The weather stations were located in areas where reliable measurements and observations could be performed in accordance with the requirements of the World Meteorological Organization (WMO) (\cite{guiwmo}). During the considered ten years period (2000~--~2009) no station was relocated. 

For comparison purposes, we have included in the present study, meteorological data recorded by weather stations deployed at the Nordic Optical Telescope (NOT) site. NOT is located at the Observatorio del Roque de Los Muchachos (ORM), in a region considered a reference for very good astronomical conditions - the Canary Islands, Spain. At ORM, NOT is positioned in the proximity of MAGIC. The NOT meteorological data have been kindly made available to the public on the web page http://www.not.iac.es/weather/. 

In order to check for regional variations and long term trends we have analyzed meteorological data on relative humidity, dew point temperature, air temperature, wind speed, barometric air pressure, and sky cloudiness. The sky cloudiness parameter has been investigated only for the Romanian locations as there is no data acquired following a similar procedure in the case of NOT.

The data records were available in the NMA database at regular intervals of one hour, similar to the case presented in (Jabiri et al. \cite{jabiri}). As the NOT data were available at a higher frequency, for a proper comparison, we have first computed hourly averages. 

Nighttime (20:00~--~06:00), daytime (06:00~--~20:00) and entire day (00:00~--~24:00) intervals have been defined and all the data series are to be considered in local time. Monthly and annual samples have been extracted from these series of raw data according to a procedure to be described in the following. 

For a particular meteorological parameter (e.g. relative humidity), location (e.g. Baisoara), month (e.g. January) and time interval (e.g. nighttime) we have extracted data from the NMA database hour-by-hour, over the ten years period. The relevant statistics (median, standard deviation, skewness) have been computed for the obtained sample distribution. Subsequently, we have repeated the procedure for each of the 12 months and the obtained statistics have been used to produce a monthly distribution of medians. The procedure employed for the generation of the annual distributions of medians is similar, but instead considering a certain month, data from a particular year (e.g. 2000) have been used. For the case of NOT, we have followed the same procedure, but hourly averages have been computed first.  

In addition to monthly and annual distributions of medians, we have also derived monthly and annual distributions of percentages. This time, for a particular parameter, location, month and time interval, we have computed the percentage of data records whose values exceed certain limits (e.g. relative humidity $>\,90\%$). The procedure has been repeated for each month and the 12 resulted values have provided a monthly distribution of percentages. The annual distributions of percentages have been obtained in a similar way.    

Within the univariate analysis performed, two statistical characteristics of each sample have been looked into: central tendency and dispersion. 

As measures of central tendency, arithmetic mean and median are usually reported. Arithmetic mean is a useful statistic only if the investigated distribution is symmetrical (not too much skew), there are no outliers, and the analyzed data are measured at interval or ratio level. Median (the middle score in a set of scores) is useful when data are not symmetrically distributed or they are measured at an ordinal level. We have always reported the median statistic for the total sky cloudiness data which have been measured at ordinal level. In what concerns the other meteorological parameters we have had to decide what the best statistic would be as a measure of central tendency. For each sample the amount of skewness has been computed using the formula:
\begin{equation}\label{eq3}
G_{1} = \frac{\sqrt{n(n-1)}}{n-2} \cdot \frac{m_{3}}{{m_{2}}^\frac{3}{2}},
\end{equation}
where $m_{3} = \frac{1}{n} \sum_{i=1}^n \left(x_{i}-\bar{x}\right)^3$ is the sample third central moment, $m_{2} = \frac{1}{n} \sum_{i=1}^n \left(x_{i}-\bar{x}\right)^2$ is the sample variance and $\bar{x}$ is the sample arithmetic mean (Joanes $\&$ Gill \cite{joanes}). However, the sample skewness tends to differ from the population skewness by a systematic amount that depends on the size of the sample. In order to be able to tell, from the samples that we have, if the populations are likely skewed (positively/negatively) or, if they are rather symmetrical, a test statistic is needed:
\begin{equation}\label{eq4}
Z = \frac{G_{1}}{SES},
\end{equation} 
where SES is the standard error of skewness given by
\begin{equation}\label{eq5}
SES = \sqrt{\frac{6n\left(n-1\right)}{\left(n-2\right)\left(n+1\right)\left(n+3\right)}}.
\end{equation}   
A population is considered to be very likely skewed if $\left|Z\right|\,>\,2$ (Cramer \cite{cramer}). Otherwise no conclusion can be reached about the skewness of that population. It might be symmetric or skewed in either direction. 

We have computed this test statistic for all the samples used in the analysis. For most of the cases it has turned out that the populations have been very likely skewed. Therefore we have chosen to use as a measure of central tendency for all the distributions involved in our analysis, the median statistic. This option has no negative impact on those cases where the test statistic given by Equation~(\ref{eq4}), did not provide a clear conclusion. If the skewness is small, the difference between the median and the arithmetic mean is also small. For a symmetric distribution, the arithmetic mean and the median are equal. 

As an indicator for the dispersion of data, we have used the standard deviation, defined as the square root of the sample variance $m_{2}$. 

Following these considerations, the relevant statistics for all the data samples used in our analysis have been the median and the standard deviation. Whenever a plot in the present paper includes error bars, the points represent the median values of the data samples for those specific months or years, while the error bars illustrate the standard deviation of the samples. As the error bars frequently overlap, in order to have a clearer image of variability in the data, we have also included bottom plots of the standard deviation values.

\section{Astroclimatological results}\label{4}

\subsection{Relative humidity}\label{4.1}

The relative humidity (\emph{RH}) has been measured at the Romanian locations to accuracy of $\pm2\%$ by sensors placed 2 m above the soil surface. At NOT, the relative humidity sensors placed 2 m above the ground provided data to accuracy better than $\pm2\%$ (Lombardi et al. \cite{lombardi2}).

\begin{figure*}
\includegraphics[width=17cm]{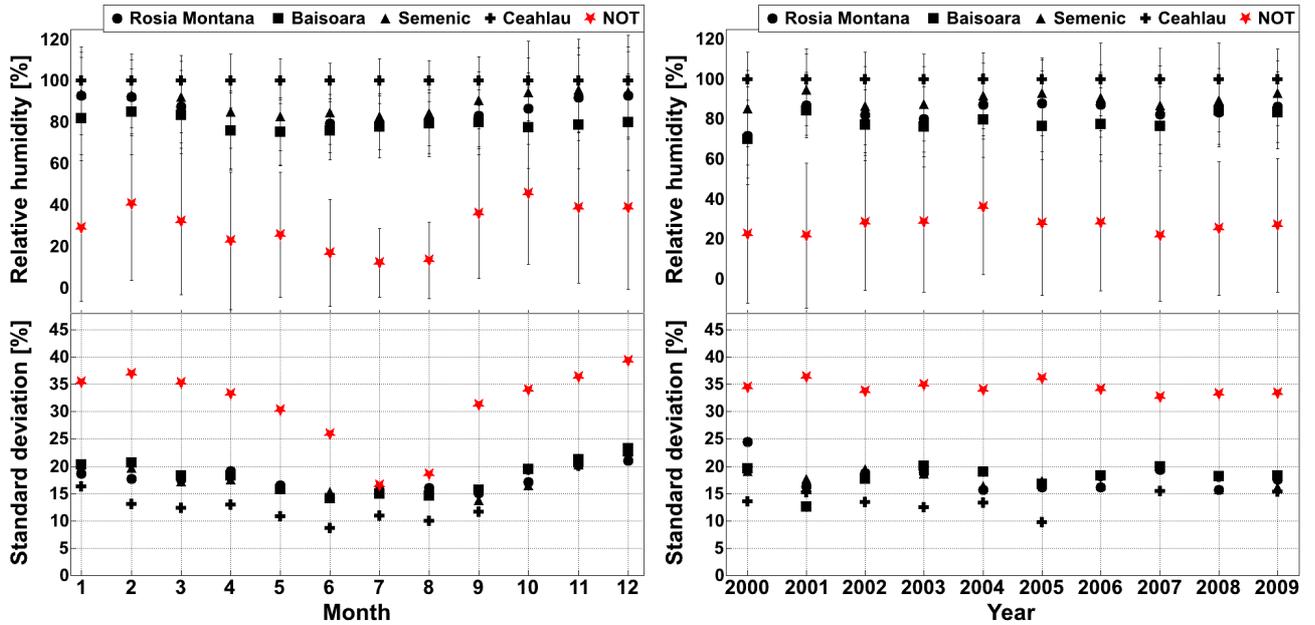}
\centering
\caption{The monthly (\emph{top-left}) and the annual (\emph{top-right}) nighttime distributions of medians for relative humidity. The bottom plots (\emph{left and right}) show the values of standard deviations.}
\label{fig1}
\end{figure*}

In Figure~\ref{fig1}, we present the monthly and the annual nighttime distributions of medians for relative humidity at the assessed sites. The bottom plots include the values of the standard deviations for each month, and respectively for each year sample. 

The medians derived for the Romanian locations are clustered together and have values more than 40$\%$ higher in comparison to NOT. The driest Romanian site is Baisoara while Ceahlau is an extreme, unfavorable case displaying relative humidity median values of 100$\%$ around the year (Figure~\ref{fig1}, \emph{top-left and right}). 

A seasonal dependence can be observed for the NOT data (Figure~\ref{fig1}, \emph{top-left}) and the lowest \emph{RH} levels are recorded in July and August. A much fainter seasonal variation can be also observed for Rosia Montana and for Semenic. Baisoara exhibits an almost flat trend over the year, but from January to March, increased levels of relative humidity can be observed.
  
The spread of data recorded at NOT exhibits a clear seasonal behavior (Figure~\ref{fig1}, \emph{bottom-left}) and it is larger than that of Romanian data (Figure~\ref{fig1}, \emph{bottom-left and right}). The minimum spread of NOT data is observed in July and August when the \emph{RH} levels also reach a minimum (Figure~\ref{fig1}, \emph{top-left}). The spread of Romanian data shows a less visible seasonal variation with a minimum also observed during summer. 

The analysis of the annual distributions of medians shows that all sites exhibit flat trends over the years with fluctuations within the statistical errors (Figure~\ref{fig1}, \emph{top-right}). For all the Romanian locations but Ceahlau, 2000 appears as the driest year of the available data. 

\begin{figure*}
\centering
\includegraphics[width=17cm]{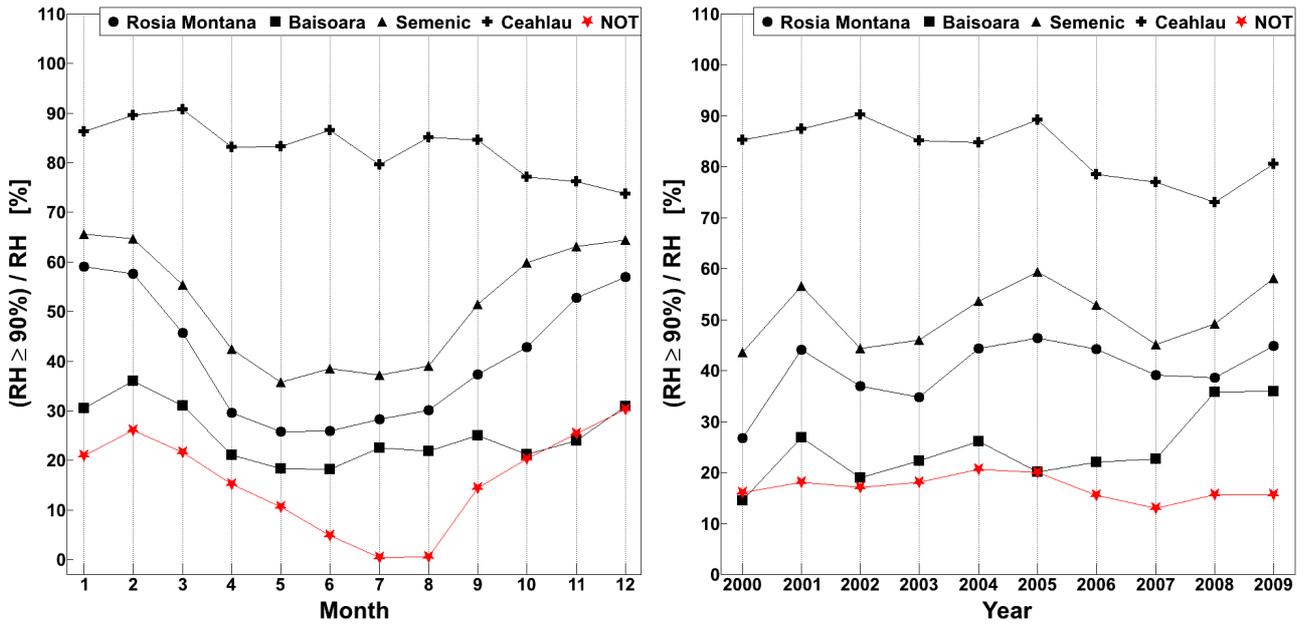}
\caption{The monthly (\emph{left}) and the annual (\emph{right}) nighttime distributions of percentages for relative humidity data records whose values are larger than 90$\%$.}
\label{fig2}
\end{figure*}

In Figure~\ref{fig2}, we display monthly and annual nighttime distributions of percentages for relative humidity data records whose values are larger than 90$\%$. Considering this value as a safety operational limit, the points in this figure indicate the percentage of time that a telescope built at the considered locations would have been unusable because of high levels of \emph{RH}. The larger are these percentages the worse are the observational conditions. A seasonal variation is visible for all locations (Figure~\ref{fig2}, \emph{left}) and the summer months appear to be best suited for observations everywhere but at Ceahlau where a small improvement can be observed only from October to December. The smallest computed values belong to NOT and they confirm that its site is excellent for astronomical observations. Among the investigated Romanian locations, Baisoara offers the closest conditions to those of NOT, especially during October, November and December.

In what concerns the annual nighttime distributions of percentages for \emph{RH} (Figure~\ref{fig2} – \emph{right}), Baisoara shows close values to NOT, but an increasing trend can be observed since 2006. This raises the question if we are in the presence of a change in the local microclimate or it is a typical oscillating behavior of the relative humidity over the years. More data are necessary to settle the issue.      

\begin{figure*}
\centering
\includegraphics[width=17cm]{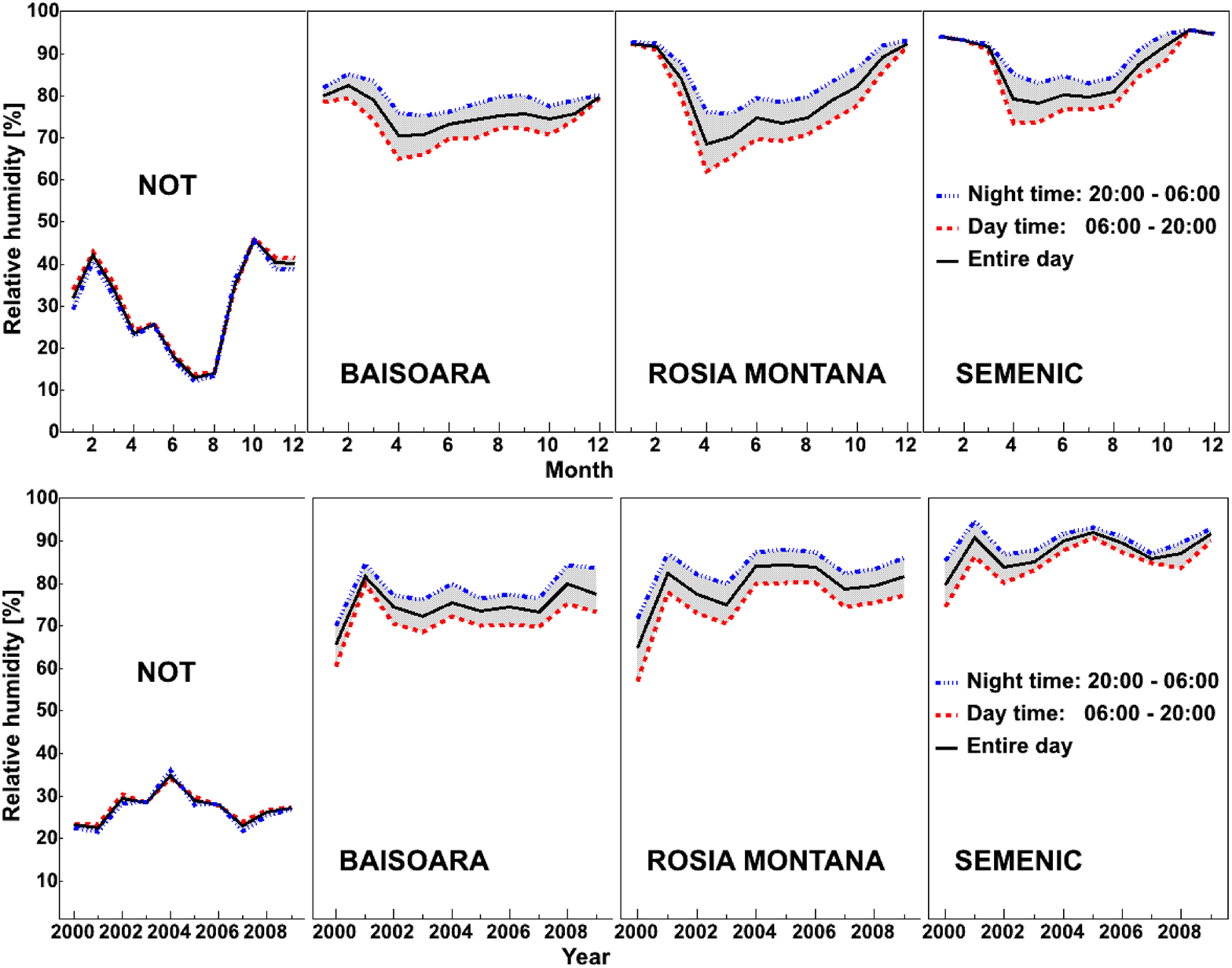}
\caption{The monthly (\emph{left}) and the annual (\emph{right}) nighttime, daytime and entire day distributions of medians for relative humidity.}
\label{fig3}
\end{figure*}

Figure~\ref{fig3} presents the monthly and the annual nighttime, daytime and entire day distributions of medians for relative humidity at Baisoara, Rosia Montana, Semenic and NOT. The median values of data recorded at Ceahlau are constantly equal to 100$\%$ and not included in this figure. The nighttime median values are higher than the daytime ones (Figure~\ref{fig3}, \emph{top and bottom}) and night-day variations of $\sim$ 20$\%$ can be observed especially during summer (Figure~\ref{fig3}, \emph{top}) at the three Romanian locations. The lower levels of \emph{RH} during daytime can be exploited for maintenance operations. At the NOT site, the night-day variations are very small and sometimes the nighttime median values are smaller than those of daytime data. This is especially favorable for the astronomical observations as they take place at night.   

\subsection{Dew point temperature}\label{4.2}

\begin{figure*}
\centering
\includegraphics[width=17cm]{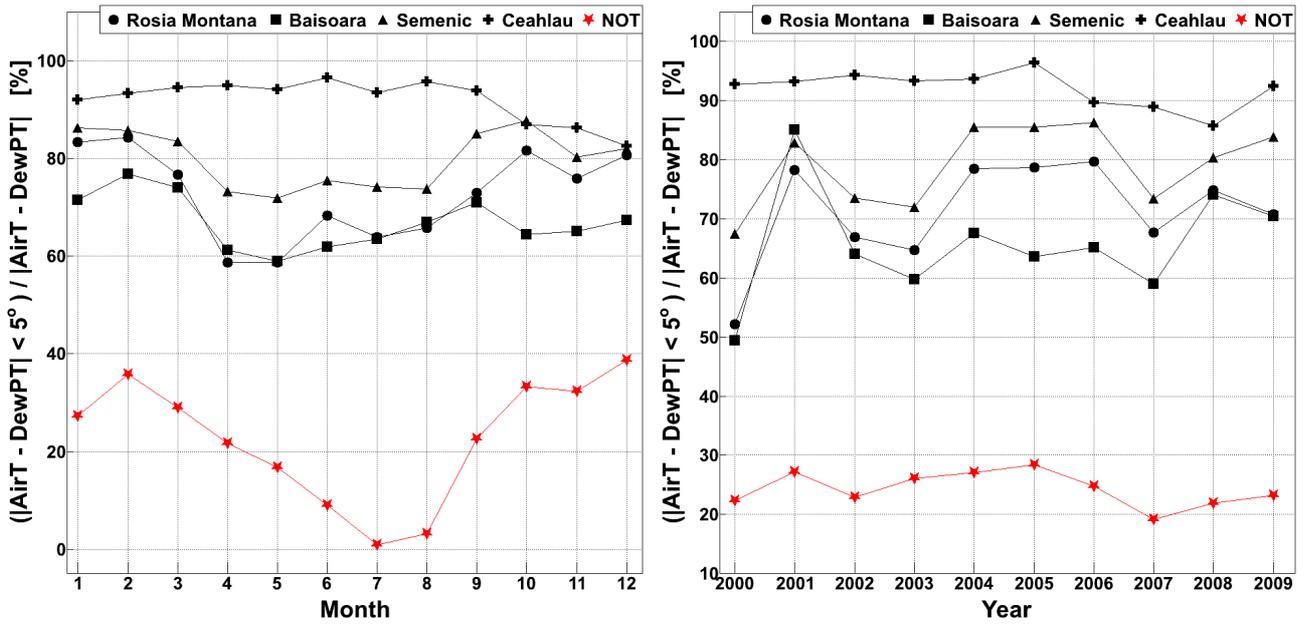}
\caption{The monthly (\emph{left}) and the annual (\emph{right}) nighttime distributions of percentages for the difference between dew point temperature and air temperature data records, whose values are smaller than $5\ ^{\circ}$C.}
\label{fig4}
\end{figure*}

In order to evaluate the risk of condensation, we have analyzed the monthly and the annual nighttime distributions of percentages for the difference between dew point temperature and air temperature data records, whose values are smaller than an upper limit. The larger are the values of these percentages for a certain site the higher is the risk of condensation expected for that particular site. Two values have been considered in our study for the upper limit: $1\ ^{\circ}$C and $5\ ^{\circ}$C. For the analyzed meteorological data of all five locations, the cases where this difference is smaller than $1\ ^{\circ}$C also imply relative humidity values larger than 90$\%$. As this situation has been analyzed in Subsection~$\S$\ref{4.1}, we present here only the case when the upper limit is set to $5\ ^{\circ}$C (Figure~\ref{fig4}). This can be regarded as an extension of the analysis performed in Subsection~$\S$\ref{4.1} to values of \emph{RH} lower than 90$\%$.
   
The location characterized by the lowest risk of condensation is that of NOT. The statistical distribution of data taken at its site shows a clear seasonal dependence (Figure~\ref{fig4}, \emph{left}) with a minimum reached in July and August. A seasonal dependence is also visible for Rosia Montana and Semenic with lower values reached during summer. Baisoara exhibits the lowest risk of condensation during April and May, while high values have been observed from January to March. Ceahlau is again the worst case even if for the interval October - December a decrease for the risk of condensation can be noticed.

The analysis of the annual distributions (Figure~\ref{fig4}, \emph{right}) shows a dry climate at NOT and confirms Baisoara as the driest Romanian location. The Romanian sites exhibit an oscillating behavior over the years and 2000 appears as the least humid one. 

\subsection{Wind speed}\label{4.3}

At the Romanian locations, wind speed has been measured by sensors placed 10 m above the soil surface, to accuracy that degrades as speed increases. The NOT wind sensor took data to accuracy better than 2$\%$ (Lombardi et al. \cite{lombardi2}).

\begin{figure*}
\centering
\includegraphics[width=17cm]{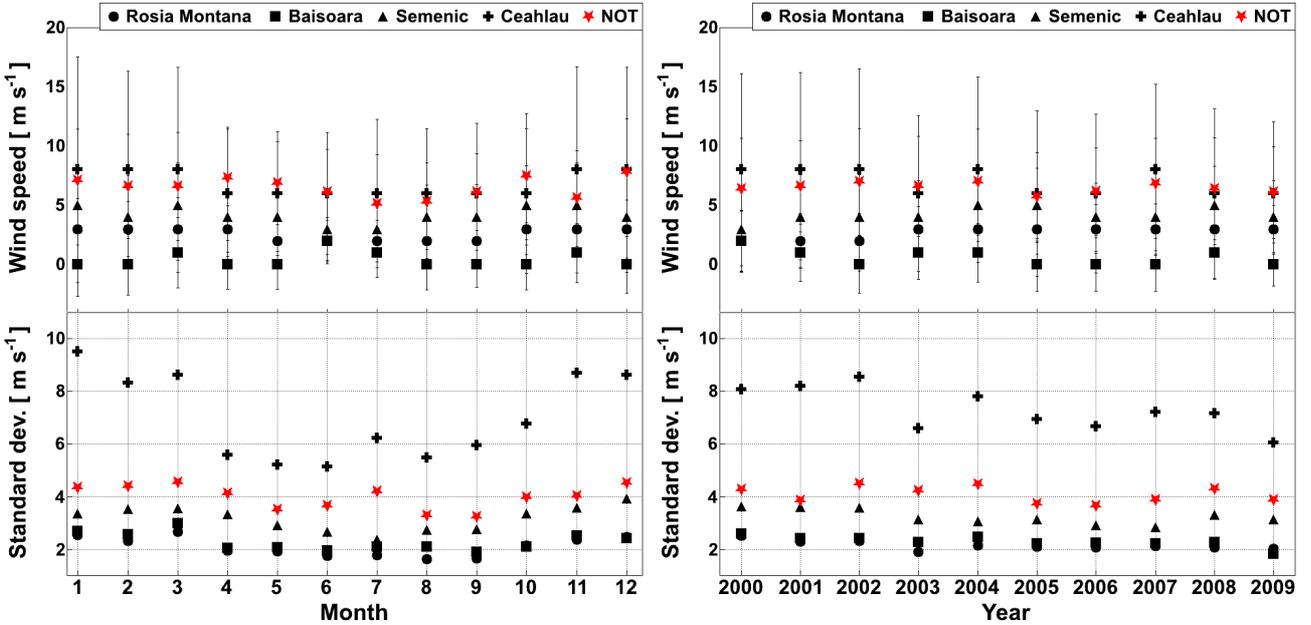}
\caption{The monthly (\emph{top-left}) and the annual (\emph{top-right}) nighttime distributions of medians for wind speed. The bottom plots (\emph{left and right}) show the values of standard deviations.}
\label{fig5}
\end{figure*}

In Figure~\ref{fig5}, we show the monthly and the annual nighttime distributions of medians for wind speed at the assessed sites. A clear dependence on altitude (\emph{see also Table~\ref{tabel1}}) can be noticed for the median values and for the spread of data. The higher is the altitude of a location, the stronger are the wind speeds and the larger are the variations of data. Baisoara is a favorable exception as its site has the lowest wind speeds even if it is not located at the lowest altitude. This effect is probably due to the local orography. The unfavorable exception is Ceahlau which has similar wind conditions to NOT and a larger spread in data, even if it is situated lower than that. 

Faint seasonal variations can be observed for the distributions derived at NOT, Ceahlau, and Semenic (Figure~\ref{fig5}, \emph{top-left}) and the lowest wind speeds are measured during summer. A seasonal variation is also visible for the spread of all Romanian data (Figure~\ref{fig5}, \emph{bottom-left}). This dependence is clearly observable at Ceahlau, but it is weaker for the other Romanian locations.

The analysis of the annual distributions (Figure~\ref{fig5}, \emph{top-right}), shows that all sites exhibit almost flat trends over the years with fluctuations within the statistical errors.  

\begin{figure*}
\centering
\includegraphics[width=17cm]{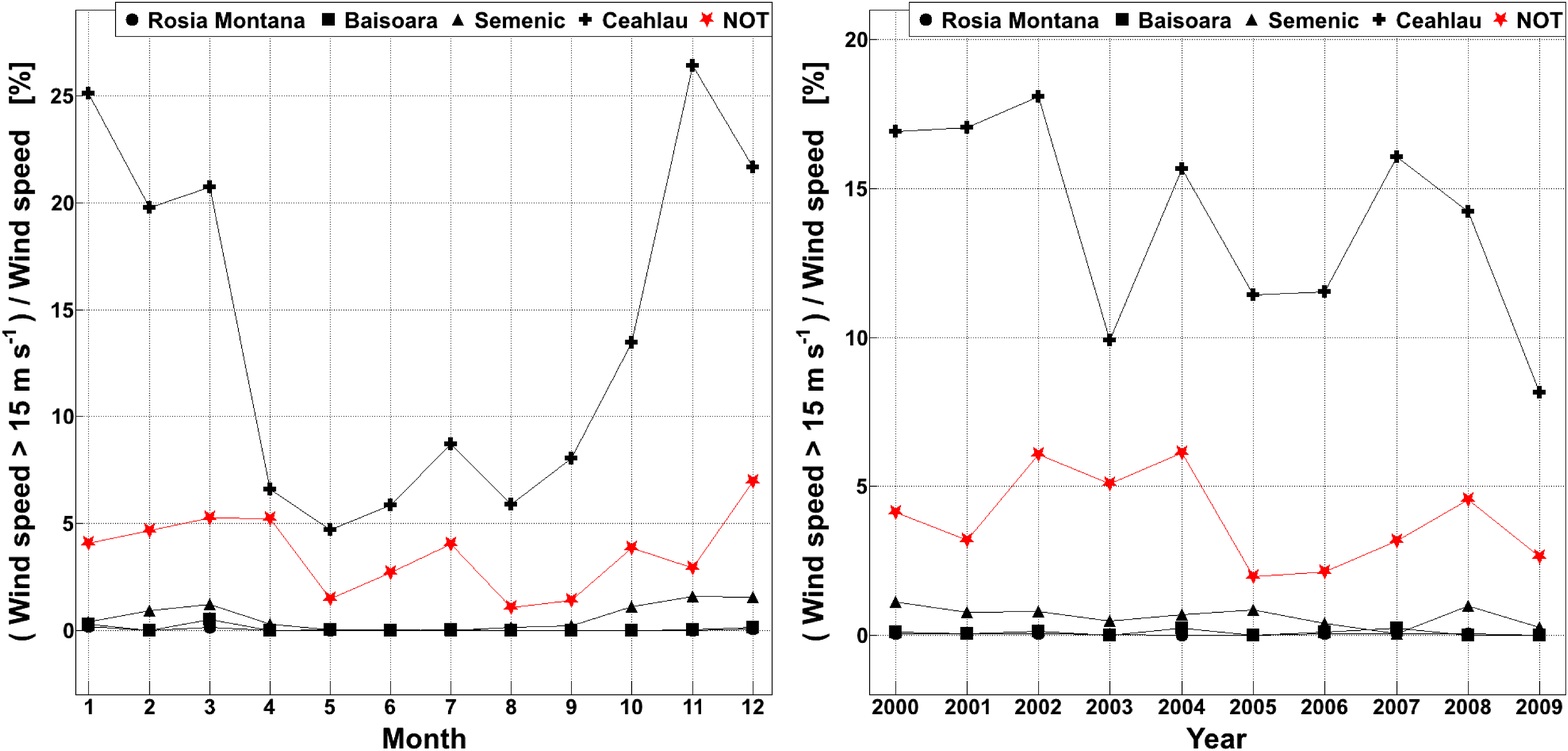}
\caption{The monthly (\emph{left}) and the annual (\emph{right}) nighttime distributions of percentages for wind speed data records whose values are larger than 15 m $s^{-1}$. }
\label{fig6}
\end{figure*}

Considering 15 m $s^{-1}$ as the wind speed limit beyond which a telescope should be turned off and brought to the parking position ($\S$\ref{2.2}), we have plotted in Figure~\ref{fig6} the monthly and the annual nighttime distributions of percentages for wind speed data records larger than this limit. The larger are these values for a certain site the higher are the percentages of time when a telescope operated there should be turned off because of high winds. All the Romanian locations, but Ceahlau show values lower than the NOT site and almost flat trends. The distribution of Ceahlau data shows a seasonal dependence (Figure~\ref{fig6}, \emph{left}) with the lowest values from April to September when its conditions are closer to those at NOT. 

A good picture of the wind speed distributions at the analyzed locations can be obtained from Table~\ref{tabel2}. It reports for the period April - October, over the ten years range, the percentages of nighttime wind speed data records whose values fit into fixed intervals. Baisoara has a predominance of wind speeds smaller than 3 m $s^{-1}$ (71.5$\%$) whereas the other locations behave worse. However, this parameter exceeds 10 m $s^{-1}$ only in 0.4$\%$ of cases at Rosia Montana and respectively 5.8$\%$ at Semenic, so these locations also preserve good wind speed conditions.

   \begin{table*}
   \caption{Nighttime wind speed statistics at Baisoara (BA), Rosia Montana (RM), Semenic (SE), Ceahlau (CE) and NOT, collected in a ten years time period (2000~--~2009), from April to October.}
   \label{tabel2}
   \centering
   \begin{tabular}{c c c c c c}
   \hline\hline
   $W_{sp}$ range [m $s^{-1}$] & BA [$\%$]	& RM [$\%$]	& SE [$\%$]	& CE [$\%$] & NOT [$\%$] \\
   \hline
   $0\,\leq\,W_{sp}\,<3$ & 71.5 & 52.6 & 28.9 & 23.5 & 16.5 \\
   $3\,\leq\,W_{sp}\,<10$ & 28.0 & 47.0 & 65.0 & 51.1 & 65.3 \\
   $10\,\leq\,W_{sp}\,\leq\,15$ & 0.5 & 0.4 & 5.8 & 17.9 & 15.4  \\
   $W_{sp}\,>15$ & 0.0 & 0.0 & 0.3 & 7.5 & 2.8  \\
   \hline
   \end{tabular}
  \end{table*}

In case 15 m $s^{-1}$ is considered as a safety operational limit, a telescope installed either at Baisoara, Rosia Montana, or Semenic will practically never have to be turned off because of strong winds.
 
If in addition to this wind speed limit we also consider that the relative humidity values should not exceed 90$\%$, a duty-cycle can be computed for each analyzed location. We have calculated this parameter only with nighttime data recorded from April to October over the ten years period and the results are presented in Table~\ref{tabel3}. 

   \begin{table}
   \caption{Nighttime duty-cycles computed at Baisoara (BA), Rosia Montana (RM), Semenic (SE), Ceahlau (CE) and NOT, as the percentage of nighttime data records with \emph{RH} $<$ 90$\%$ and wind speed $\leq$ 15 m $s^{-1}$, from April to October (2000~--~2009).}
   \label{tabel3}
   \centering
   \begin{tabular}{c c c c c c}
   \hline\hline
    & BA & RM & SE	& CE & NOT \\
   \hline
   Duty-cycle [$\%$] & 78.9 & 68.6 & 56.4 & 16.5 & 87.9 \\
   \hline
   \end{tabular}
   \end{table}

The best result for a Romanian location has been obtained for Baisoara that would offer a duty-cycle of $\sim$ 79$\%$. This value is quite close to the one obtained for the reference site of NOT ($\sim$ 88$\%$) and it recommends Baisoara as a good potential site for the establishment of a Cherenkov telescope in Romania.

\subsection{Air temperature}\label{4.4}

The air temperature at the Romanian locations was measured by sensors placed 2 m above the soil surface to accuracy better than $\pm0.2\ ^{\circ}$C. The same type of measurements was performed at NOT to accuracy of $\pm0.1\ ^{\circ}$C.

\begin{figure*}
\centering
\includegraphics[width=17cm]{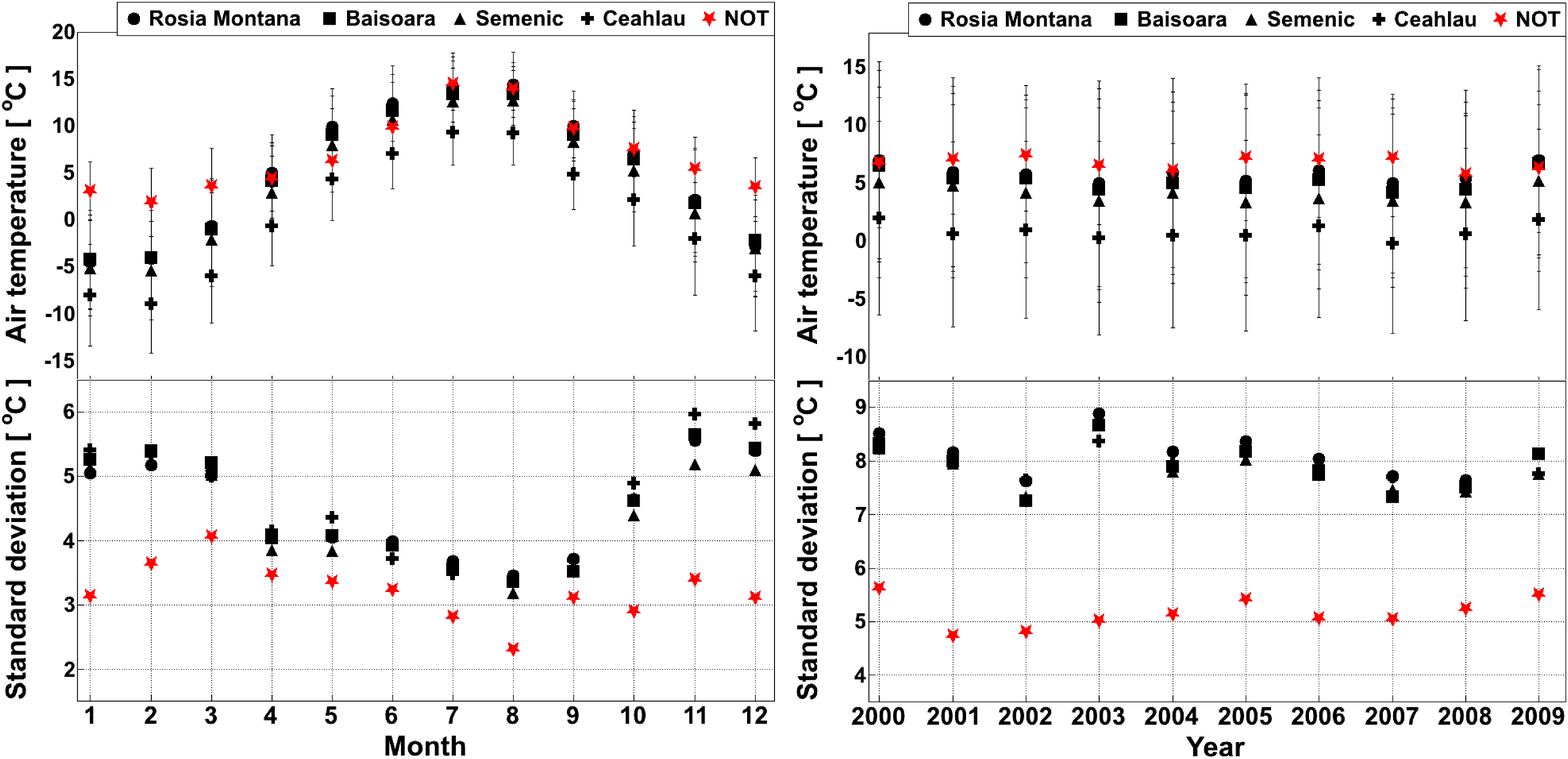}
\caption{The monthly (\emph{top-left}) and the annual (\emph{top-right}) nighttime distributions of medians for air temperature. The bottom plots (\emph{left and right}) show the values of standard deviations.}
\label{fig7}
\end{figure*}

In Figure~\ref{fig7}, we display the monthly and the annual nighttime distributions of medians for air temperature. A clear seasonal variation is noticeable for all investigated sites, but the difference between winter and summer months is higher for the Romanian locations (Figure~\ref{fig7}, \emph{top-left}). Ceahlau is characterized by the lowest air temperatures around the year, while the other Romanian locations exhibit similar median values. From April to October the air temperature conditions at Baisoara, Rosia Montana and Semenic are very similar to those at NOT.

A larger spread characterizes air temperature data at the Romanian locations in comparison to NOT (Figure~\ref{fig7} \emph{bottom-left and right}) and a clear seasonal dependence can be observed from Figure~\ref{fig7} (\emph{bottom-left}). Less variability in the Romanian air temperature data occurs from April to September when its values tend to get closer to those of the NOT site. 

The analysis of the annual distributions (Figure~\ref{fig7}, \emph{top-right}) shows for all sites almost flat trends with fluctuations within the statistical errors. 

Figure~\ref{fig7} (\emph{top-left and right}) also depicts an altitude dependence (\emph{see also Table~\ref{tabel1}}) for the median values of air temperature distributions. Ceahlau, the highest altitude of the investigated Romanian locations is also the coldest one. As the altitude decreases, the temperature values increase. Rosia Montana and Baisoara exhibit the highest temperatures. Even if located at the highest altitude, NOT is also characterized by high temperatures. This effect can be explained by the much lower latitude where NOT is positioned in comparison to the Romanian locations (Table~\ref{tabel1}). The increase in altitude is compensated by the decrease in latitude and the median temperatures increase.   

\begin{figure*}
\centering
\includegraphics[width=17cm]{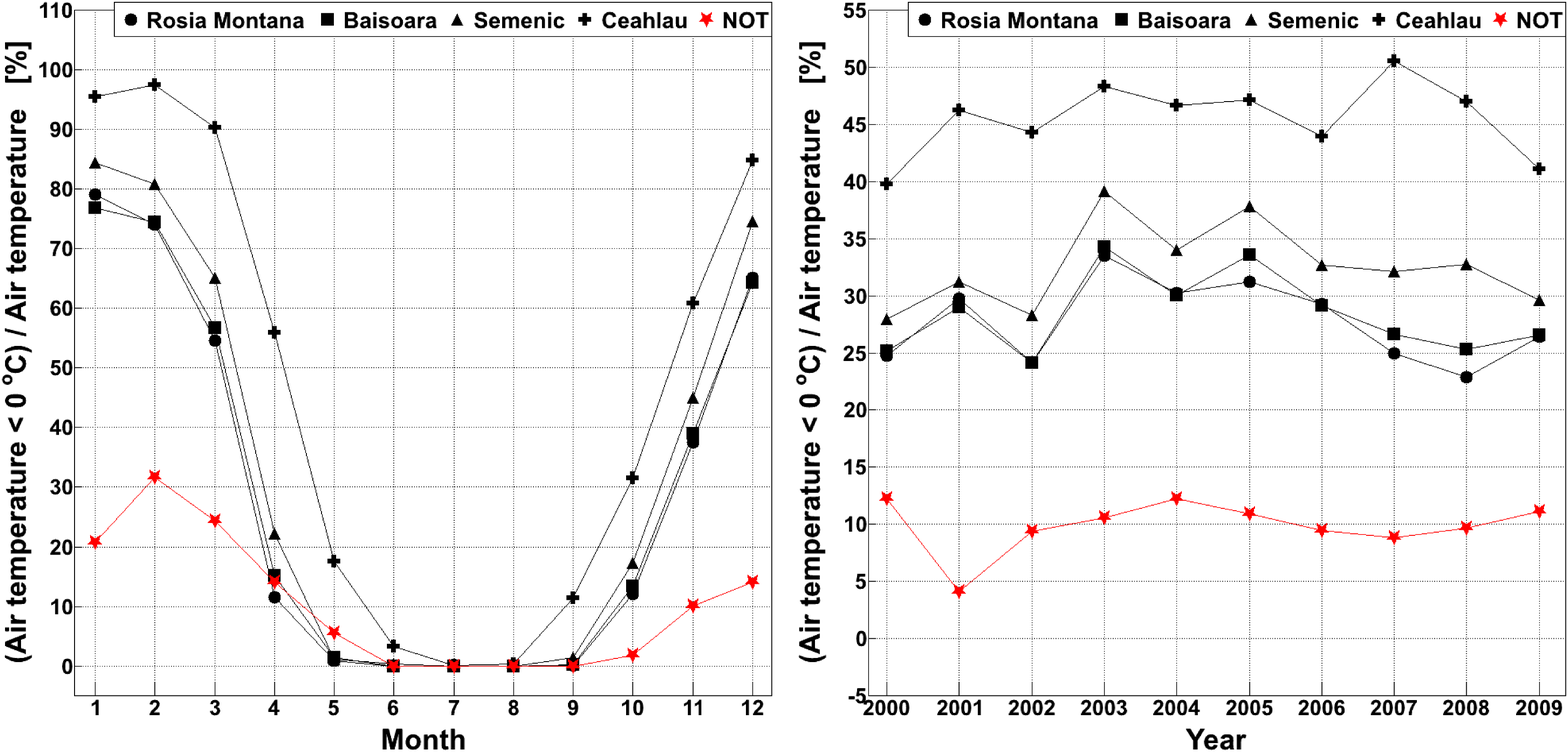}
\caption{The monthly (\emph{left}) and the annual (\emph{right}) nighttime distributions of percentages for air temperatures data records whose values are below $0\ ^{\circ}$C.}
\label{fig8}
\end{figure*}

Special technical procedures can be employed to overcome the negative effect of freezing on the proper functioning of telescopes. However, an ideal site to host a telescope would very rarely have air temperatures below $0\ ^{\circ}$C. Figure~\ref{fig8} shows the monthly and the annual nighttime distributions of percentages for air temperatures data records whose values are below $0\ ^{\circ}$C. The larger are these values for a certain site the higher are the periods of time when a telescope operated there would be affected by freezing. From the analysis of these distributions results that in Romania, the best period for observations is May - September (Figure~\ref{fig8}, \emph{left}) when data are quite similar to those at NOT and there is practically no freezing. The annual distributions (Figure~\ref{fig8}, \emph{right}) show larger year-to-year fluctuations at the Romanian sites in comparison to NOT (exception for 2000 and 2001). Baisoara and Rosia Montana experience the least freezing among the Romanian sites. 

\begin{figure*}
\centering
\includegraphics[width=17cm]{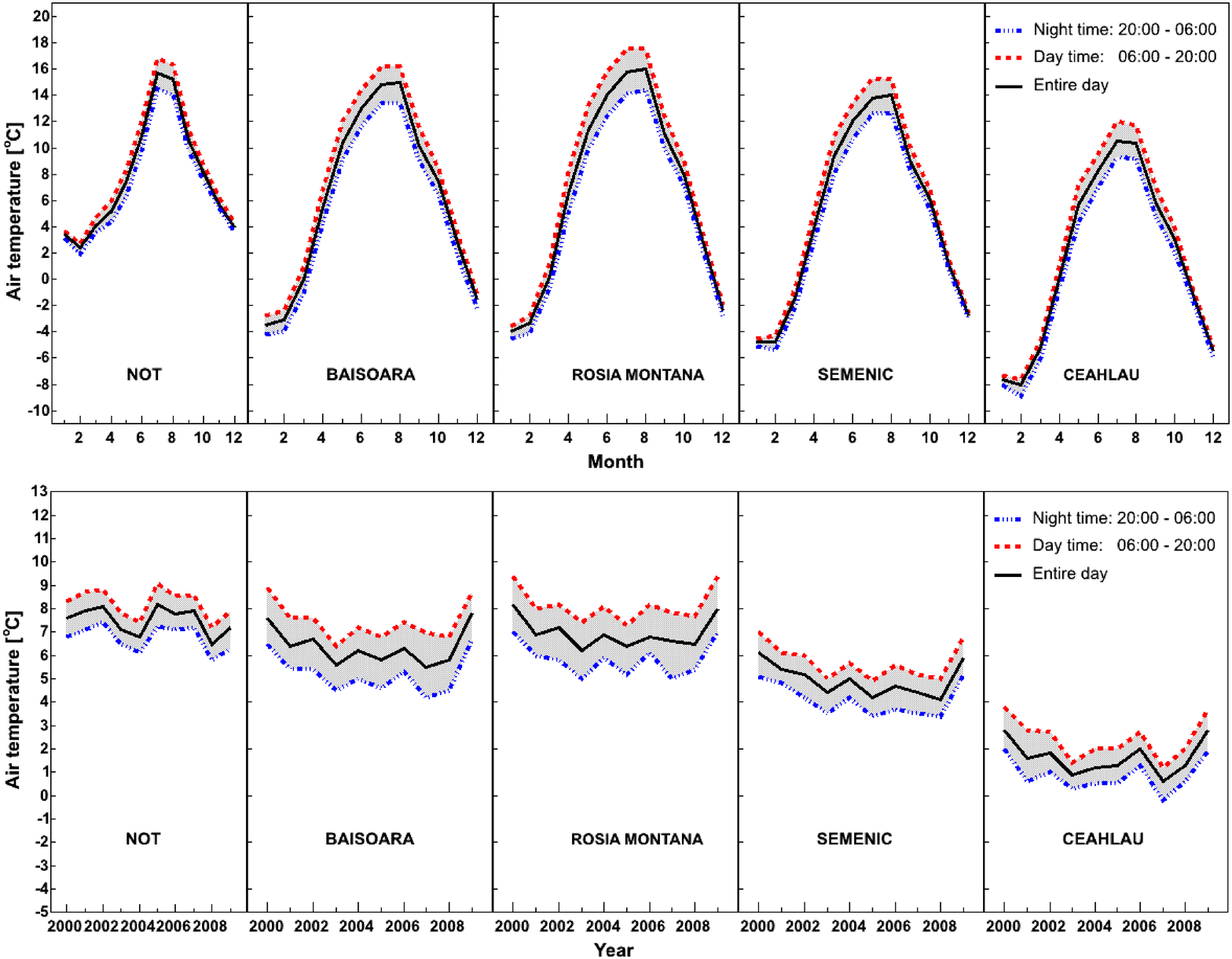}
\caption{The monthly (\emph{top}) and the annual (\emph{bottom}) nighttime, daytime and entire day distributions of medians for air temperature.}
\label{fig9}
\end{figure*}

The median values of the monthly and the annual nighttime, daytime, and entire day distributions of air temperature are presented in Figure~\ref{fig9}. For all locations, the largest daytime to nighttime variations can be observed during summer (Figure~\ref{fig9}, \emph{top}) and larger differences can be noticed at the Romanian locations compared to NOT. The annual distributions (Figure~\ref{fig9}, \emph{bottom}) show similar levels for the daytime temperatures at Baisoara, Rosia Montana and NOT, while the nighttime temperatures are lower in Romania. 

\subsection{Barometric air pressure}\label{4.5}

The barometric air pressure was measured at the Romanian locations 1.20 m above the soil surface to accuracy of $\pm$0.1 hPa. The NOT air pressure sensor was placed 2 m above the ground and it provided data to accuracy of $\pm$0.1 hPa (Lombardi et al. \cite{lombardi2}).  

\begin{figure*}
\centering
\includegraphics[width=17cm]{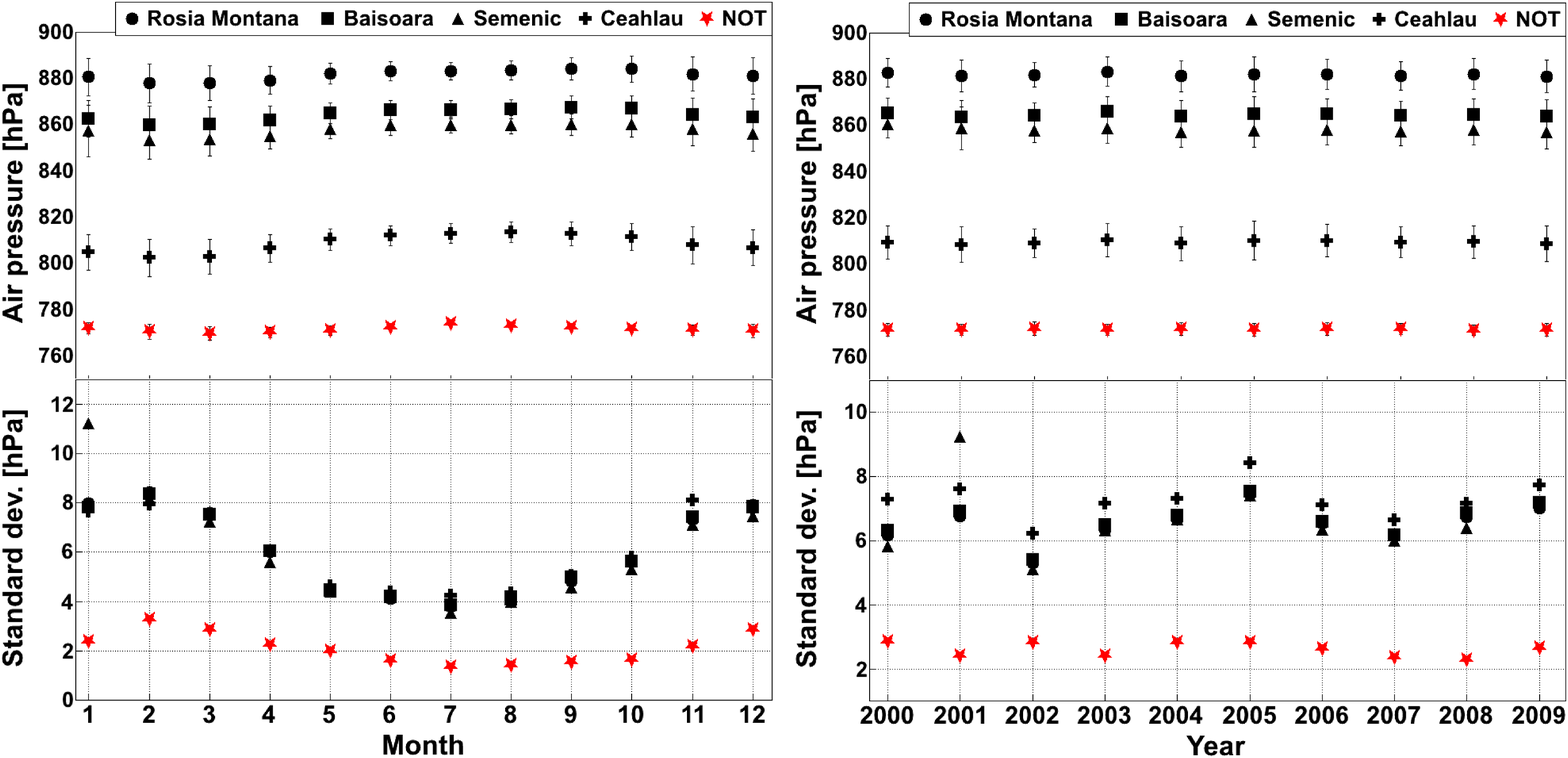}
\caption{The monthly (\emph{top-left}) and the annual (\emph{top-right}) nighttime distributions of medians for barometric air pressure. The bottom plots (\emph{left and right}) show the values of standard deviations.}
\label{fig10}
\end{figure*}

Figure~\ref{fig10} shows the monthly and the annual nighttime distributions of medians for barometric air pressure at the studied sites. A clear inverse correlation with altitude can be observed in the \emph{top-left} and the \emph{top-right} sections of the figure (\emph{see also Table~\ref{tabel1}}). The highest altitude site (NOT) exhibits the lowest air pressures, whereas the highest pressures have been recorded at the lowest altitude site (Rosia Montana).
 
A very faint seasonal variation can be observed from Figure~\ref{fig10} (\emph{top-left}) at all Romanian locations with higher median values of barometric air pressure recorded during summer. The analysis of the annual distributions (Figure~\ref{fig10}, \emph{top-right}) shows an almost constant flat trend over the years for all locations.

The spread of data recorded at the Romanian sites shows a more pronounced seasonal dependence and it is larger in comparison to NOT (Figure~\ref{fig10}, \emph{bottom-left}). The lowest spread of data is observed during summer when the weather conditions are more stable and the air pressure values fluctuate less. The lower spread of NOT air pressure data is also an indicator of a more stable weather.    

\begin{figure*}
\centering
\includegraphics[width=17cm]{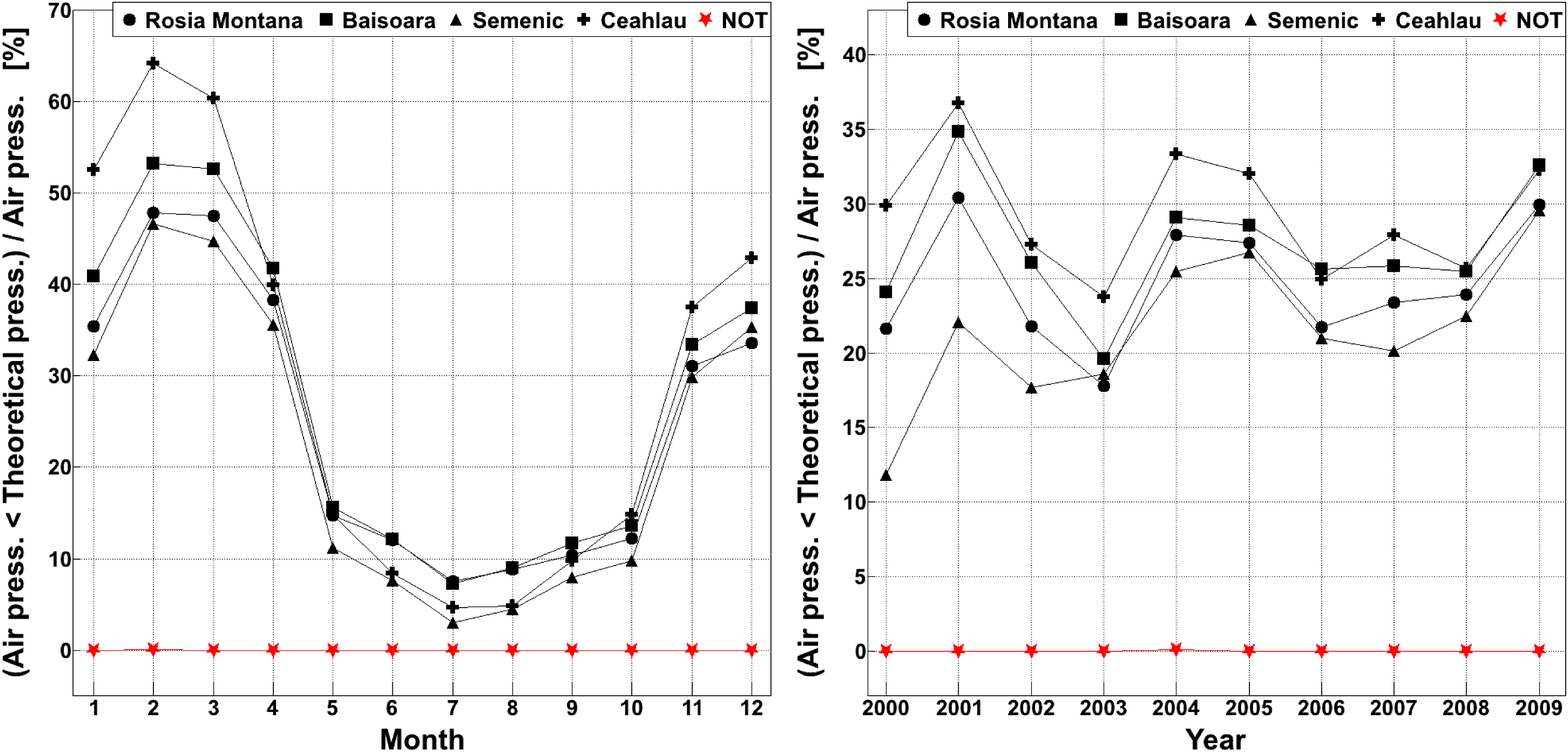}
\caption{The monthly (\emph{left}) and the annual (\emph{right}) nighttime distributions of percentages for local barometric air pressure data records whose values are lower than the theoretically expected barometric air pressure values given in Table~\ref{tabel1}.}
\label{fig11}
\end{figure*}

When the local barometric air pressure is lower than the theoretical value ($\mathrm{P_{theo}}$) at the same altitude, the weather conditions may become unstable ($\S$\ref{2.4}). In order to compare the stability of weather conditions, we have computed for the investigated sites the percentages of local barometric air pressure data records whose values are lower than those theoretically expected. The lower are these percentages for a certain site, the better are the observational conditions offered by that particular site. The results are presented in Figure~\ref{fig11}. 

The percentage values at NOT are roughly equal to 0$\%$. Almost always the values of local barometric air pressure are higher than those theoretically expected. The site is dominated by high pressure values and remarkable stable weather conditions characterize it. The situation is different at the Romanian sites where the values of the local barometric air pressure often fall below those theoretically expected. Figure~\ref{fig11} (\emph{left}) shows a very clear seasonal dependence with the most favorable weather conditions for astronomical observations from May to October.

The analysis of the annual distributions (Figure~\ref{fig11}, \emph{right}) shows that for all Romanian sites these percentages fluctuate extensively. Large differences can be noticed for the years 2000, 2001, 2002, 2004, and 2007, while smaller ones have appeared in 2003, 2005, 2006, 2008, and 2009. 

From the distributions depicted in Figure~\ref{fig11}, Semenic appears to have the most stable weather of the Romanian investigated sites. 

\subsection{Sky cloudiness}\label{4.6}

For our analysis we have used a parameter stored in the database of the NMA and called ``total sky cloudiness''. This is a parameter characteristic for the sky condition and it represents the degree to which the sky is covered by any type of clouds at the moment of observation. For the Romanian locations and the period of time considered (2000~--~2009) this parameter was visually determined by specially trained human observers. They went out on meteorological platforms and from a spot that allows maximum visibility of the celestial vault, evaluated the total sky cloudiness on a scale that stretched from 0 to 9. ``Zero'' meant a ``perfectly clear sky''. ``Eight'' stood for a ``completely covered sky'' and ``nine'' was a ``sky invisible because of fog''.
         
\begin{figure*}
\centering
\includegraphics[width=17cm]{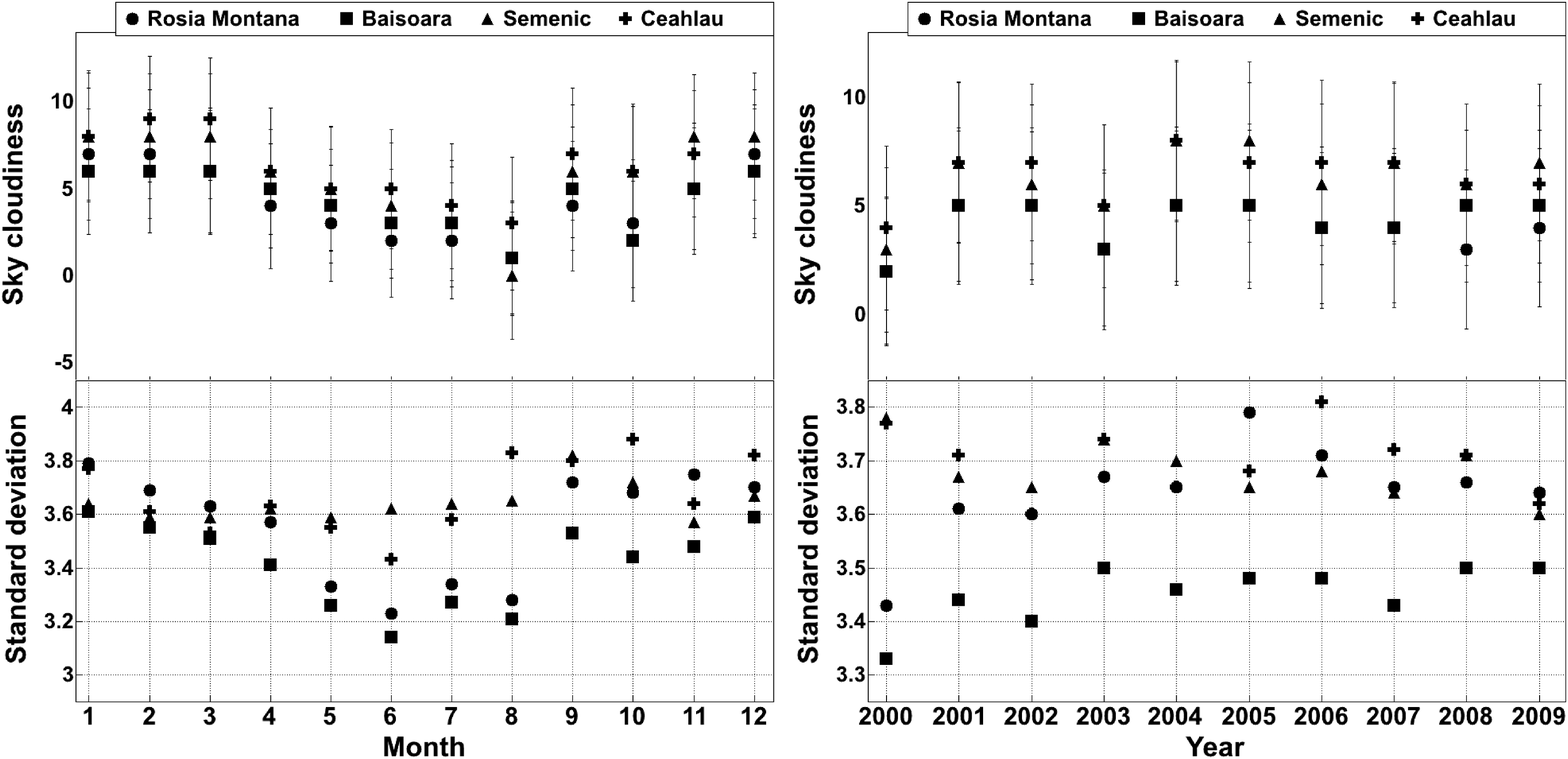}
\caption{The monthly (\emph{top-left}) and the annual (\emph{top-right}) nighttime distributions of medians for total sky cloudiness. The bottom plots (\emph{left and right}) show the values of standard deviations.}
\label{fig12}
\end{figure*}

Figure~\ref{fig12} exposes the monthly and the annual nighttime distributions of medians for total sky cloudiness at the Romanian sites. The monthly nighttime distributions of total sky cloudiness (Figure~\ref{fig12}, \emph{top-left}) show a seasonal variation with the lowest cloudiness during summer. Rosia Montana and Baisoara offer the lowest levels of cloudiness and August appears to be the best month for observations. The analysis of the annual distributions (Figure~\ref{fig12}, \emph{top-right}) confirms the quality of these two locations.

Baisoara and Rosia Montana show a seasonal dependence for its spread of data with the lowest values obtained during summer (Figure~\ref{fig12}, \emph{bottom-left}). The standard deviations included in the plot of the annual distributions (Figure~\ref{fig12}, \emph{bottom-right}) show the lowest spread for the data collected at Baisoara.

The levels of total sky cloudiness data recorded in the NMA database do not reflect the thickness of the cloud layers and implicitly the quality of the atmospheric transmission. A level eight sky is a completely covered sky, but the layer of clouds can be a thin one and the VHE gamma-ray observations might still be possible. On the other hand, under a level five sky (partly covered), thick clouds may frequently move into the telescope's field of view and it may seriously impede on taking high quality data. 

\begin{figure*}
\centering
\includegraphics[width=17cm]{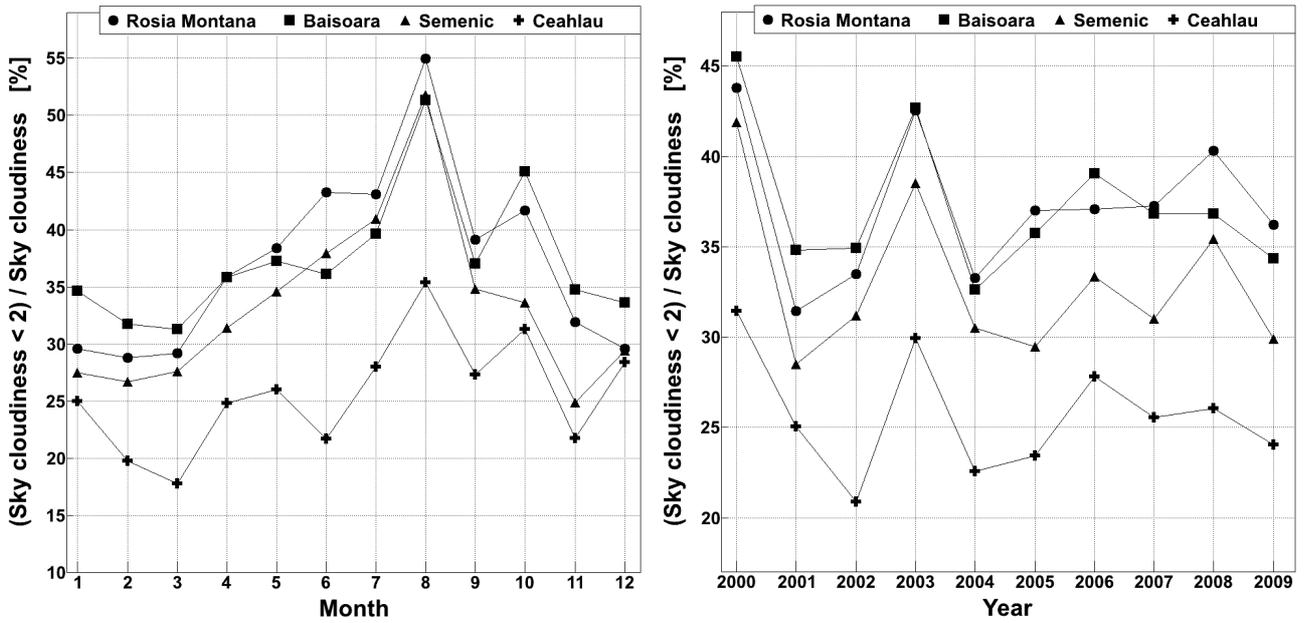}
\caption{The monthly (\emph{left}) and the annual (\emph{right}) nighttime distributions of percentages for total sky cloudiness data records whose values are strictly smaller than 2.}
\label{fig13}
\end{figure*}

Under these circumstances we have computed the percentages of total sky cloudiness data records whose values are strictly smaller than 2. A sky of level 0 or 1 is perfectly clear and it will certainly allow high quality observations of very high energy gamma-rays. The higher are these values for a certain site the better are the observational conditions offered there. The results are presented in Figure~\ref{fig13} (\emph{left and right}) and they show that Baisoara and Rosia Montana exhibit the best observational conditions, while August is the month when sky is most frequently clear.

\begin{figure*}
\centering
\includegraphics[width=17cm]{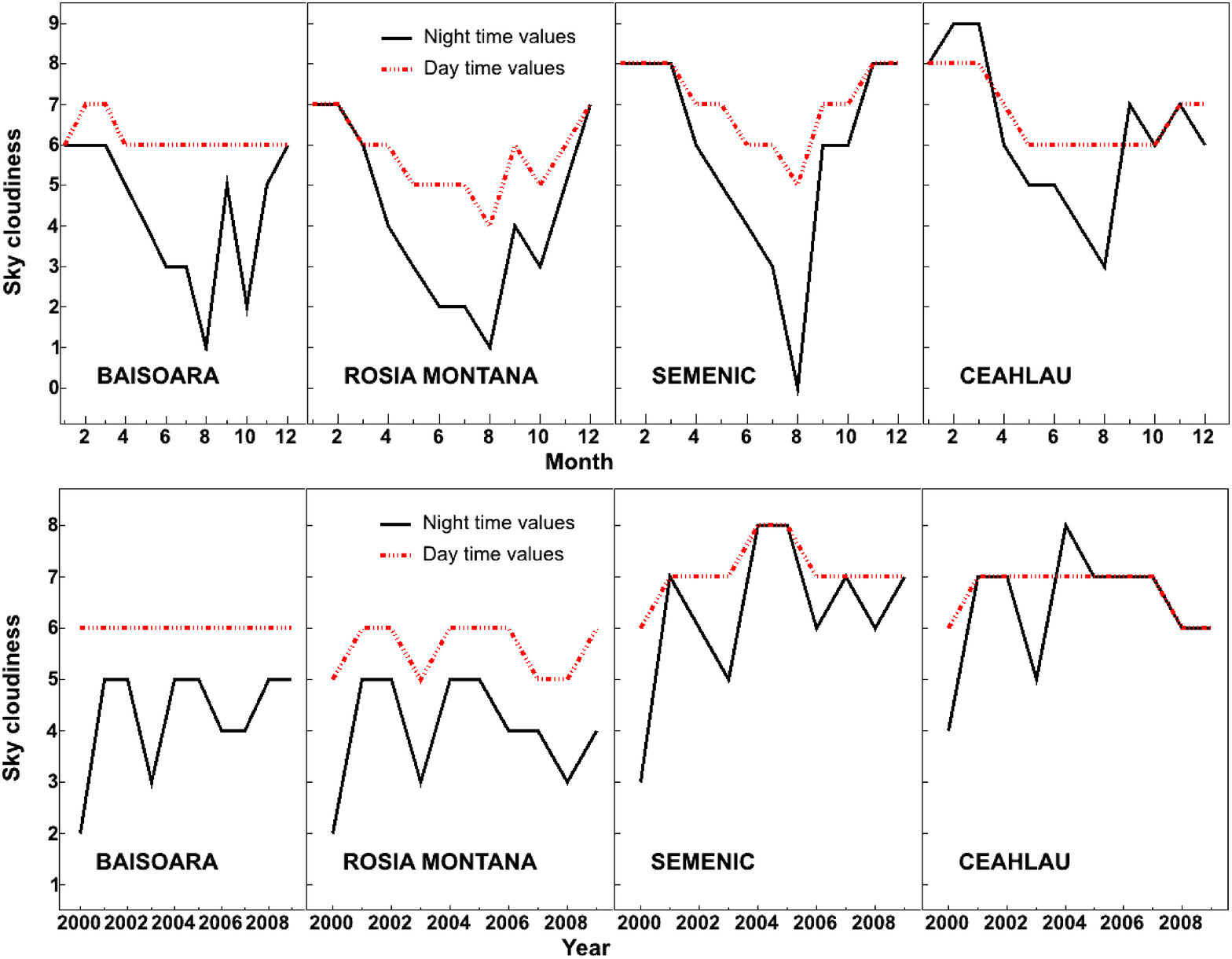}
\caption{The monthly (\emph{top}) and the annual (\emph{bottom}) nighttime and daytime distributions of medians for total sky cloudiness.}
\label{fig14}
\end{figure*}

In Figure~\ref{fig14}, we have plotted the monthly and the annual nighttime and daytime distributions of medians for total sky cloudiness. At Baisoara, the total sky cloudiness is lower during nighttime than during daytime, around the year. This is a particularly favorable effect for the astronomical observations as they take place at night. For the other locations (Figure~\ref{fig14}, \emph{top}), the effect is observable only on certain periods (e.g. March~--~September at Ceahlau). The annual distributions (Figure~\ref{fig14}, \emph{bottom}) show nighttime lower than daytime values of total sky cloudiness at Baisoara and Rosia Montana over the analyzed ten years period. 

\section{Conclusions}
A site assessment campaign has been initiated for the identification of the best location where to install and operate a small, imaging atmospheric Cherenkov telescope in Romania. 

The results of an astroclimatological study based on the statistical analysis of relative humidity, dew point temperature, air temperature, wind speed, barometric air pressure, and sky cloudiness local data have been presented in this paper. 

Data over a ten years period (2000~--~2009) have been sampled from the archive of the National Meteorological Administration (NMA) of Romania, for four candidate locations (Baisoara, Rosia Montana, Semenic, Ceahlau). Meteorological data recorded at the Nordic Optical Telescope site (Canary Islands) have been used as a reference for comparison purposes. 

Prior to presenting the results of the analysis, the influence of atmospheric conditions on the operation of Cherenkov telescopes has been reviewed and the statistical characteristics of the available meteorological data have been studied in order to choose the relevant statistics of distributions.  

The analysis of relative humidity data has revealed, as expected that the weather in Romania is more humid than at NOT. However, among the Romanian investigated sites, the microclimate of Baisoara appears to be the driest one and it offers the closest conditions to those reported for NOT. During summer the relative humidity nighttime data exhibit the least spread and they are clustered around low level central values at all investigated locations, but Ceahlau. Night-day variations of up to $\sim$ 20$\%$ have been observed between the medians of the statistical distributions at the Romanian sites, while at NOT these variations are very small. 

In order to evaluate the risk of condensation, we have analyzed the nighttime distributions of percentages for the difference between dew point temperature and air temperature data records, whose values are smaller than $5\ ^{\circ}$C. The lowest risk of condensation is characteristic to NOT, while in Romania, Baisoara offers the best conditions. 

The wind speed analysis has shown that the higher is the altitude of a site, the stronger are the winds and the larger is the spread of data. Baisoara is an exception being characterized by the lowest winds, even if it is not located at the lowest altitude. Speeds smaller than 3 m $\mathrm{s^{-1}}$ have been measured there in 71.5$\%$ of the time. Wind speed conditions are generally better in Romania than at NOT especially due to the lower altitudes.

A duty-cycle has been computed considering that a telescope should be turned off in case wind speed exceeds 15 m $\mathrm{s^{-1}}$ or the relative humidity is larger than 90$\%$. For samples including nighttime data recorded from April to October over the ten-years period, the best result of the Romanian locations has been obtained for Baisoara ($\sim$ 79$\%$). This value is quite close to that obtained for the reference site of NOT ($\sim$ 88$\%$).

Very similar air temperature conditions have been noticed at Baisoara, Rosia Montana and Semenic in comparison to NOT for the period April to October and practically no freezing occurs at these locations from May to September. An inverse correlation has been observed between air temperature and altitude. Rosia Montana and Baisoara exhibit the highest temperatures, while Ceahlau is the coldest one. Even if located at a higher altitude than Ceahlau, NOT exhibits high temperatures. This effect can be explained by the much lower latitude where NOT is positioned in comparison to the Romanian locations. Slightly larger night-day variations of air temperature have been detected in Romania during summer compared to NOT.

The analysis of the barometric air pressure data has pointed out that the NOT site is dominated by high pressure values and remarkably stable weather conditions characterize it. At the Romanian sites the weather conditions are more stable from May to October and Semenic has been designated as the location with the most stable weather. The inverse correlation between the barometric air pressure and the altitude has been observed for all locations. 

Data of total sky cloudiness have been available only for the Romanian locations and our study has indicated that Baisoara and Rosia Montana offer the largest percentages of time when the sky is perfectly clear. The analysis of the night-day variations has allowed us to observe that at Baisoara the nighttime cloudiness is lower on average than at daytime. This effect is particularly favorable for the astronomical observations.

The results of the present study have made us select Baisoara as the best investigated site for the establishment of a Cherenkov telescope in Romania. Good meteorological conditions have been also determined at Rosia Montana and as these two sites are both located in the Apuseni Mountains, we consider this area as the optimal place for the location of the telescope. A suitable period, around the year, for operating a Cherenkov telescope in Romania, appears to be from April to October, but taking data is not restricted to this time frame only. However, beyond this period the number of usable nights is lower.

\begin{acknowledgements}

The authors are grateful to Wolfgang Rhode and Michael Backes for the very enlightening discussions. This work was supported by the Romanian Ministry of Education, Research, Youth and Sport, the National Center for the Management of Programs (CNMP), through grant 81-010/2007/Program 4. The results published in this paper are partly based on data provided by the Nordic Optical Telescope, operated on the island of La Palma jointly by Denmark, Finland, Iceland, Norway, and Sweden, in the Spanish Observatorio del Roque de los Muchachos of the Instituto de Astrofisica de Canarias.

\end{acknowledgements}

\end{document}